\documentstyle[12pt]{article}
\newcommand{\be}{\begin{equation}}
\newcommand{\bea}{\begin{eqnarray}}
\newcommand{\eea}{\end{eqnarray}}
\newcommand{\ba}{\begin{array}}
\newcommand{\ea}{\end{array}}
\newcommand{\ee}{\end{equation}}

\expandafter\ifx\csname mathbbm\endcsname\relax

\else

\fi
\begin{document}

\begin{titlepage}
 \hfill \vbox{
    \halign{#\hfil         \cr
           hep-th/0406204  \cr
           } 
      }  
\vspace*{20mm}
\begin{center}
{\Large {\bf On the Seiberg-Witten map of ${\cal N}=2$ SYM theory
in Non(anti)commutative Harmonic Superspace }}

 \vspace*{15mm}
\vspace*{1mm} {Batool Safarzadeh \footnote{BSAFAR@modares.ac.ir}} \\
\vspace*{1cm}

{\it Department of Physics, School of Sciences \\
Tarbiat Modares University, P.O. Box 14155-4838, Tehran, Iran\\
}

\vspace*{1cm}
\end{center}

\begin{abstract}
We consider ${\cal N}=2$ supersymmetric $U(1)$ gauge theory in a
nonanticommutative ${\cal N}=2$ harmonic superspace with the
singlet deformation. We generalize analytic superfield and gauge
parameter to the nonanticommutative theory so that gauge
transformations act on the component fields in a canonical
form (Seiberg-Witten map). This superfield,
upon a field redefinition transforms under supersymmetry in a standard way.
\end{abstract}
\vspace{1cm}
\end{titlepage}

The deformation of superspace has been studied intensively
\cite{Ooguri:2003qp},\cite{Ferrara:2000mm},\cite{Klemm:2001yu},\cite{deBoer:2003dn},
\cite{ferrara2000qp}.
 Bosonic \cite{Seiberg:1999vs} and fermionic \cite{Seiberg:2003yz}
deformations of superspace have given rise to
non-commutative ([x,x]=0) and non(anti)commutative
$(\{\theta,\theta\}\neq0)$ coordinates respectively. In the latter case, for
${\cal N}=1$, half of supersymmetry is broken and the
unbroken Q supersymmetry is known as ${\cal N}=\frac{1}{2}$
\cite{Seiberg:2003yz}. It is an interesting problem
to study the deformation of the extended superspace $({\cal
N}=2)$. Depending on whether one chooses the supercovariant
derivatives $D_{\alpha}$ or the supersymmetry generators
$Q_{\alpha}$ as  the defferential operators defining the Poisson
brackets, one obtains the full ${\cal N}=2$ supersymmetry
\cite{Ferrara:2003xk} or a partial ${\cal N}=2$ supersymmetry
\cite{Ivanov:2003te},\cite{Araki:2004mq} respectively.
 The singlet deformation of ${\cal N}=2$
supersymmetric $U(1)$ gauge theory in the harmonic superspace
which breaks half of ${\cal N}=2$ supersymmetry has been studied in
\cite{Ferrara:2004zv},\cite{Araki:2004de}. In this case, the gauge
and supersymmetric transformations get corrections which are linear in the
deformation parameter. It is interesting to generalize analytic
superfield and gauge parameter to the non(anti)commutative theory so
that gauge transformations of the component fields have canonical
forms (Seiberg-Witten map). Seiberg and Witten claimed in
\cite{Seiberg:1999vs} that certain noncommutative gauge theories
are equivalent to commutative ones. In particular, they argued
that there exists a map from a commutative gauge field to a
noncommutative one which is compatible with the gauge structure
of each.
 In this work we
will apply this idea to  $U(1)$ ${\cal N}=2$ with a singlet deformation and obtain
the deformed supersymmetry transformations of
 the component fields . Finally we will redefine component fields
such that the standard form of supersymmetric transformations is restored and
then use them to find the action which turns out to have a simple form.

We begin by introducing the non(anti)commutative deformation of
${\cal N}=2$ harmonic superspace. ${\cal N}=2$ in the four
dimensional
superspace is parameterized by\\
$(x^{\mu},\theta^{\alpha},\bar{\theta}^{ \dot{\alpha}},u^{+i})$
where $\mu=0,1,2,3$, $\alpha, \dot{\alpha}=1,2$ and $i=1,2$ are
spacetime, spinor and  $SU(2)_{R}$ indices respevtively. In this
harmonic superspace, supersymmetry generators $Q_{\alpha}^{\pm},
\bar{Q}_{ \dot{\alpha}\pm}$ and supercovariant $D_{\alpha}^{\pm}
, \bar{D}_{\alpha}^{\pm}$ derivatives are defined by
\bea\label{rrr} Q_{\alpha}^{\pm}&=&u_{i}^{\pm}Q_{\alpha}^{i},
\;\;\bar{Q}_{\alpha}^{\pm}=u_{i}^{\pm}\bar{Q}_{\alpha}^{i},\cr
&&\cr D_{\alpha}^{\pm}&=&u_{i}^{\pm}D_{\alpha}^{i},
\;\;\bar{D}_{\alpha}^{\pm}=u_{i}^{\pm} \bar{D}_{\alpha}^{i}, \eea
and analytic superfields are defined by the differential
constraint \bea
 D_{\alpha}^{+}\phi=
\bar{D}_{\alpha}^{+}\phi(x_{A}^{\mu},\theta^{+},
\bar{\theta}^{+},u)=0.\eea

The non(anti)commutativity in the ${\cal N}=2$ harmonic superspace
is introduced by Moyal-Weyl star product
 \bea \{\theta _{i}^{\alpha},\theta
_{j}^{\beta}\}_{*}=\frac{1}{4}\epsilon
_{ij}\epsilon^{\alpha\beta}C_{s}. \eea Here the *-product is
defined by \bea
f(\theta)*g(\theta)&=&f(\theta)exp(P)g(\theta),\cr  &&\cr
P&=&-\frac{1}{8}\epsilon_{ij}\epsilon^{\alpha\beta}C_{s}
\overleftarrow{Q}_{\alpha}^{i}\overrightarrow{Q}_{\beta} ^{j}\;. \eea

The action of ${\cal N}=2$ supersymmetric $U(1)$ gauge theory in
this non(anti)commutative harmonic superspace is written in terms
of an analytic superfield $V^{++}$\cite{Zupnik:vm} \bea
S=\frac{1}{2}\sum_{n}^{\infty}\frac{(-i)^n}{n}\int d^{4}x
d^{8}\theta du_{1}\ldots
du_{n}\frac{V^{++}(\zeta_{1},u_{1})*\ldots
*V^{++}(\zeta_{n},u_{n})}
{(u_{1}^{+}u_{2}^{+})\ldots(u_{n}^{+}u_{1}^{+})}\;,\eea where
$\zeta_{i}=(x_{A},\theta_{i}^{+}, \bar{\theta}_{i}^{+})$ and $
d^{8}\theta=d^{4}\theta^{+}d^{4}\theta^{-}$. The action is
invariant under the gauge transformation \bea \label{drv}
\delta_{\Lambda}^{*}V^{++}&=&-D^{++}\Lambda+i[\Lambda,V^{++}]_*\;.
\eea

The gauge parameter $\Lambda(\zeta,u)$ is also analytic. $D^{++}$
harmonic derivative is defined by \bea
D^{++}=u^{+i}\frac{\partial}{\partial u^{-i}}-2i\theta^{+}
\sigma^\mu\bar{\theta}^{+}\frac{\partial}{\partial x_A^{\mu}}
+\theta^{+\alpha}\frac{\partial}{\partial\theta^{-\alpha}}
+\bar{\theta}^{+\dot{\alpha}}\frac{\partial}{\partial
\bar{\theta}^{- \dot{\alpha}}}\;. \eea

Since the analytic superfild and gauge parameter contain an
infinite number of (auxilary) fields in the harmonic superspace
representation, the gauge freedom can be used to set some of their
components to zero \cite{Galperin2000}. In the Wess-Zumino(WZ)
gauge we then find \bea
V_{WZ}^{++}(\xi,u)=&-&i\surd{2}(\theta^{+})^{2}\bar{\phi}+
i\surd{2}(\bar{\theta}^{+})^{2}\phi-
2i(\theta^{+}\sigma^\mu\bar{\theta}^{+})A_{\mu}\cr &&\cr
&+&4(\bar{\theta}^{+})^{2}\theta^{+}\psi^{i}u_{i}^{-}
-4(\theta^{+})^{2} \bar{\theta}^{+}\bar{\psi}^{i}u_{i}^{-}\\ &&\cr
&+&3(\theta^{+})^{2}(\bar{\theta}^{+})^{2}
D^{ij}u_{i}^{-}u_{j}^{-} ,\;\;\;\;\;\; and\;\;\;\;
 \Lambda=\lambda(x_{A}^{\mu})\nonumber.
 \eea

In the case of singlet deformation, the gauge variation of
$V_{WZ}^{++}$ is calculated as \bea
\delta_{\lambda}^{*}V_{WZ}^{++}&=&+2i(\theta^{+}\sigma^\mu\bar{\theta}^{+})
\partial_{\mu}\lambda
-\frac{1}{8}\epsilon^{\alpha\beta}\epsilon_{ij}C_{s}\bigg{[}4i\surd{2}
\theta_{\alpha}^{+}\sigma_{\beta\dot{\beta}}^{\mu}
(\bar{\theta}^{\dot{\beta}})\partial_{\mu}\lambda\bar{\phi}\cr
&&\cr
&-&4\sigma_{\alpha\dot{\alpha}}^{\mu}\sigma_{\beta\dot{\beta}}^{\nu}
\frac{i}{2}\epsilon^{\dot{\alpha}\dot{\beta}}
(\bar{\theta}^{+})^{2}\partial_ {\mu}\lambda A_{\nu}
+16\sigma_{\beta\dot{\beta}}^{\mu}(\bar{\theta}^{+})^{\dot{\beta}}
(\bar{\theta}^{\dot{\alpha}})\theta_{\alpha}^{+}
\bar{\psi}^{i}u_{i}^{-}\partial_{\mu}\lambda\bigg{]}\cr  &&\cr
&&\bigg{[}u^{+j}u^{-i}-u^{+i}u^{-j}\bigg{]}\\ &&\cr
&=&-2i(\theta^{+}\sigma^\mu\bar{\theta}^{+})
\left[-\partial_{\mu}\lambda-\frac{1}{\surd{2}}C_{s}\partial_{\mu}\lambda
\bar{\phi}\right]+i(\bar{\theta}^{+})^{2}
C_{s}\partial_{\mu}\lambda A^{\mu}\cr &&\cr
&+&4(\bar{\theta}^{+})^{2}(\theta^{+})^{\beta}\left(\frac{-1}{2}\partial_{\mu}
\lambda (\sigma^{\mu}\bar{\psi^{i}})_{\beta}
C_{s}u_{i}^{-}\right)\nonumber. \eea Deformed gauge
transformations for the component fields in the case of the
singlet deformation read as
 \bea\label{drv00}
 \delta _{\Lambda}^{*}
A_{\mu}&=&-\bigg{(}1+\frac{1}{\surd{2}}C_{s}
\bar{\phi}\bigg{)}\partial _{\mu}\lambda \;,\cr  &&\cr \delta
_{\Lambda}^{*}\phi&=&\frac{1}{\surd{2}}C_{s}A_{\mu} \partial
^{\mu}\lambda\;,\\   &&\cr
 \delta _{\Lambda}^{*}\psi
 _{\alpha}^{i}&=&-\frac{1}{2}C_{s}
\partial _{\mu}\lambda(\sigma^{\mu} \bar{\psi}^{i})_{\alpha}\;,\cr
&&\cr
 \delta _{\Lambda}^{*}\bar{\phi}&=&\delta
 _{\Lambda}^{*}\bar{\psi}_{ \dot{\alpha}}^{i}=\delta
_{\Lambda}^{*}D^{ij}=0 \nonumber. \eea

We may also consider another point of view in which we
change the definition of the superfields and the supergauge
parameter such that the gauge transformations of the component
fields get the standard form. In comparision with the previous
situation, this point of view could then be interpreted as the
Seiberg-Witten map for the non(anti)commutative ${\cal N}=2$ SYM
theory \cite{Mikulovic:2004qj}.

 To be precise let us consider a deformed analytic superfield $V_{WZ}^{++}$ as follows
 \bea\label{drv0}
V_{WZ}^{++}(\xi,u)&=&-i\surd{2}(\theta^{+})^{2}\bar{\phi}+
i\surd{2}(\bar{\theta}^{+})^{2}\phi-
2i(\theta^{+}\sigma^\mu\bar{\theta}^{+})A_{\mu}+
4(\bar{\theta}^{+})^{2}\theta^{+}\psi^{i}u_{i}^{-} \cr &&\cr
&-&\frac{i}{2}\epsilon_{ij}\epsilon^{\alpha\beta}(\bar{\theta}^{+})^{2}
\theta_{\beta}^{+}\sigma_{\alpha\dot{\alpha}}^{\mu}
\{(\bar{\psi}^{i})^{\dot{\alpha}},A_{\mu}\}u^{-j}
-4(\theta^{+})^{2} \bar{\theta}^{+}\bar{\psi}^{i}u_{i}^{-} \\
&&\cr &+&3(\theta^{+})^{2}(\bar{\theta}^{+})^{2}
\left[D^{ij}-\frac{C_{s}}{3\surd{2}}\epsilon^{ij}\partial
_{\mu}\{A^{\mu},\bar{\phi}\}\right]u_{i}^{-}u_{j}^{-} .\nonumber
\eea Accordingly we will also change the gauge parameter
$\Lambda$ as the following
 \bea
\Lambda=\lambda+\frac{i}{4}\epsilon_{ij}u^{-i}u^{-j}C_{s}
(\bar{\theta}^{+})^{2}\{\partial_{\mu}\lambda,A_{\mu}\}
+\frac{i}{2\surd{2}}\epsilon_{ij}u^{-i}u^{-j}C_{s}
(\theta^{+}\sigma^\mu\bar{\theta}^{+}) \{\partial_
{\mu}\lambda,\bar{\phi}\}. \eea

 In order to find the gauge transformation of the component fields one needs
to compute the effect of $D^{++}$ on the gauge parameter $\Lambda$
\bea \label{drv1}
 D^{++}\Lambda
&=&-2i(\theta^{+}\sigma^\mu\bar{\theta}^{+})\partial_
{\mu}\lambda+\frac{2i}{4}\epsilon_{ij}u^{+i}u^{-j}C_{s}
(\bar{\theta}^{+})^{2}\{\partial_{\mu}\lambda,A^{\mu}\}\cr &&\cr
&+&\frac{i}{\surd{2}}\epsilon_{ij}u^{+i}u^{-j}C_{s}
(\theta^{+}\sigma^\mu\bar{\theta}^{+})\{\partial_
{\mu}\lambda,\bar{\phi}\}\\ &&\cr &+&
\frac{1}{\surd{2}}\epsilon_{ij}u^{-i}u^{-j}C_{s}
(\theta^{+})^{2}(\bar{\theta}^{+})^{2}\partial ^{\mu}\{ \partial
_{\mu}\lambda,\bar{\phi}\}.\nonumber \eea
 Moreover we find
  \bea \label{drv3}
[\lambda,V_{WZ}^{++}]&=&\bigg{[}\lambda,\frac{-i}{2}C_{s}\epsilon_{ij}
\epsilon^{\alpha\beta}(\bar{\theta}^{+})^{2}
\theta_{\beta}^{+}\sigma_{\alpha\dot{\alpha}}^{\mu}
\{(\bar{\psi}^{i})^{\dot{\alpha}},A_{\mu}\}u^{-j}\bigg{]}\cr &&\cr
&=&\frac{-i}{2}\epsilon_{ij}\epsilon^{kl}\bigg{[}u_{k}^{+}u_{l}^{-},u^{-j}\bigg{]}
\epsilon^{\alpha\beta}(\bar{\theta}^{+})^{2}
\theta_{\beta}^{+}\sigma_{\alpha\dot{\alpha}}^{\mu} \lambda
\{(\bar{\psi}^{i})^{\dot{\alpha}},A_{\mu}\}\\ &&\cr
&=&-iC_{s}(\bar{\theta}^{+})^{2}
\theta^{+\alpha}\sigma_{\alpha\dot{\alpha}}^{\mu} \lambda
\{(\bar{\psi}^{i})^{\dot{\alpha}},A_{\mu}\}u_{i}^{-}.\nonumber
\eea Here we have used the fact that the gauge group is $U(1)$
and therefore all other terms in $\Lambda$ will commute with
$V^{++}$.

Plugging (\ref{drv1}) and (\ref{drv3}) into (\ref{drv}), the gauge transformation of
 the deformed analytic superfield (\ref{drv0}) reads
\bea
\delta_{\lambda}^{*}V_{WZ}^{++}&=&-2i(\theta^{+}\sigma^\mu\bar{\theta}^{+})
\bigg{[}-\partial_{\mu}\lambda-\frac{1}{\surd{2}}C_{s}\partial_{\mu}\lambda
\bar{\phi}\bigg{]}+i(\bar{\theta}^{+})^{2}
C_{s}\partial_{\mu}\lambda A^{\mu}\cr &&\cr
&+&4(\bar{\theta}^{+})^{2}(\theta^{+})^{\beta}\bigg{(}\frac{-1}{2}\partial_{\mu}\lambda
(\sigma^{\mu}\bar{\psi^{i}})_{\beta} C_{s}u_{i}^{-}\bigg{)}
-\frac{2i}{4}C_{s}
(\bar{\theta}^{+})^{2}\{\partial_{\mu}\lambda,A^{\mu}\}\\ \cr
&-&\frac{1}{\surd{2}}C_{s}
(\theta^{+})^{2}(\bar{\theta}^{+})^{2}\partial ^{\mu}\{ \partial
_{\mu}\lambda,\bar{\phi}\}u^{-i}u^{-j} -\frac{i}{\surd{2}}C_{s}
(\theta^{+}\sigma^\mu\bar{\theta}^{+})\{\partial_
{\mu}\lambda,\bar{\phi}\}\cr &&\cr
&+&C_{s}(\bar{\theta}^{+})^{2}\theta^{+\alpha}\sigma_{\alpha\dot{\alpha}}^{\mu}
\lambda
\{(\bar{\psi}^{i})^{\dot{\alpha}},A_{\mu}\}u_{i}^{-}\nonumber \eea

 From this
relation one can read the gauge transformations of the components
of the superfeild in the case of singlet deformation as follows
 \be \delta _{\Lambda}^{*} A_{\mu}=-\partial
_{\mu}\lambda ,\;\;\;\;\; \delta _{\Lambda}^{*}\phi=\delta
_{\Lambda}^{*}\psi
 _{\alpha}^{i}=\delta _{\Lambda}^{*}\bar{\phi}=\delta
 _{\Lambda}^{*}\bar{\psi}_{ \dot{\alpha}}^{i}=\delta
_{\Lambda}^{*}D^{ij}=0 .\ee
 which is the same as the ordinary
field theory. Therefore by making use of the deformed analytic
superfield (\ref{drv0}) the gauge transformations of the component
fields (\ref{drv00}) reduce to the canonical form.

   Let us now write down the corresponding
Lagrangian using the deformed superfield (\ref{drv0}). By making
use of the same method as in \cite{Araki:2004mq}, the Lagrangian
up to the first order of $(C_{s})$ can be computed. The result is
\bea L&=&\frac{1}{4}F_{\mu\nu}(F^{\mu\nu}+
\tilde{F}^{\mu\nu})-i\psi ^{i}\sigma ^{\mu}\partial
_{\mu}\bar{\psi}_{i}-\partial ^{\mu}\phi \partial _{\mu}
\bar{\phi}+\frac{1}{4}D_{ij}D^{ij}\cr &&\cr
&+&\frac{1}{\surd{2}}C_{s}A_{\nu}\partial
_{\mu}\bar{\phi}(F^{\mu\nu}+
\tilde{F}^{\mu\nu})+\frac{i}{\surd{2}}C_{s} \bar{\phi}\bigg{(}
\psi^{k}\sigma ^{\nu}\partial _{\nu}
\bar{\psi}_{k}\bigg{)}+\frac{i}{\surd{2}}C_{s}\bigg{(}
\psi^{k}\sigma ^{\nu}\bar{\psi}_{k}\bigg{)}\partial _{\nu}
\bar{\phi} \cr&& \cr &+&\frac{i}{2}C_{s}
\bar{\psi}^{i}\bar{\psi}^{j}D_{ij}+\frac{\surd{2}}{4}C_{s}A_{\mu}A^{\mu}\partial
^{2} \bar{\phi}-\frac{\surd{2}}{4}C_{s} \bar{\phi}D^{ij}D_{ij}
.\eea where
$\tilde{F}^{\mu\nu}=\frac{i}{2}\epsilon^{\mu\nu\rho\sigma}F_{\rho\sigma}$.

 The next step would be to check how the
supersymmetry transformations $\delta _{\xi}$  of the component
fields work for this deformed superfield. The supersymmetry
transformation in the WZ gauge and for the non(anti)commutative
case has been studied in \cite{Araki:2004de} which has the
following form \bea
\delta_{\xi}V_{WZ}^{++}&=&\tilde{\delta}_{\xi}V_{WZ}^{++}+\delta
_{\Lambda}^{*}V_{WZ}^{++} \cr  &&\cr
\tilde{\delta}_{\xi}V_{WZ}^{++}&\equiv& \xi_{i}^{\alpha}Q_{\alpha}
^{i}V_{WZ}^{++}, \eea
 where in the analytic basis one has
\begin{eqnarray*}
\xi_{i}^{\alpha}Q_{\alpha}
^{i}=-\xi^{+\alpha}Q_{\alpha}^{-}+\xi^{-\alpha}Q_{\alpha}^{+},
\end{eqnarray*}
with
\begin{eqnarray*}
Q_{\alpha} ^{+}&=&\frac{\partial}{\partial \theta
^{-\alpha}}-2i\sigma_{\alpha \dot{\alpha}}^{\mu} \bar{\theta}^{+
\dot{\alpha}}\frac{\partial}{\partial x_{A}^{\mu}},\cr &&\cr
Q_{\alpha}^{-}&=&-\frac{\partial}{\partial \theta^{+\alpha}}.
\end{eqnarray*}

To preserve the WZ-gauge we must use the most general analytic
gauge parameter $\Lambda (\zeta,u)$: \bea \Lambda
(\zeta,u)&=&\lambda ^{(0,0)}(x_{A},u)+\bar{\theta}_{
\dot{\alpha}}^{+}\lambda ^{(0,1) \dot{\alpha}}(x_{A},u)+\theta
^{+\alpha}\lambda _{\alpha}^{(1,0)}(x_{A},u)\cr  &&\cr
&+&(\bar{\theta}^{+})^{2}\lambda
^{(0,2)}(x_{A},u)+(\theta^{+})^{2}\lambda
^{(2,0)}(x_{A},u)+\theta^{+}\sigma^{\mu} \bar{\theta}^{+}
\lambda_{\mu}^{(1,1)}(x_{A},u)\\  &&\cr &+&
(\bar{\theta}^{+})^{2}\theta^{+\alpha} \lambda
^{(1,2)}(x_{A},u)+(\theta^{+})^{2}\bar{\theta}_{
\dot{\alpha}}^{+}\lambda ^{(2,1)
\dot{\alpha}}(x_{A},u)+(\theta^{+})^{2}(\bar{\theta}^{+})^{2}\lambda
^{(2,2)}(x_{A},u). \nonumber\eea

 Using the appropriate gauge parameter and the deformed analytic superfield
(\ref{drv0}), it is easy to see how the various fields transform
\bea
\delta_{\xi}\phi&=&-\surd{2}i\xi^{i}\psi_{i}-\frac{i}{2\surd{2}}C_{s}\xi^{i}
\sigma^{\mu}\{ \bar{\psi}_{i},A_{\mu}\}\;,\cr  &&\cr
\delta_{\xi}\bar{\phi}&=&0\;,\cr &&\cr
\delta_{\xi}A_{\mu}&=&i\xi^{i}\sigma_{\mu} \bar{\psi}_{i}\;,\cr
&&\cr
\delta_{\xi}\psi_{\alpha}^{i}&=&\bigg{(}1+\frac{1}{\surd{2}}C_{s}
\bar{\phi}\bigg{)}(\sigma^{\mu\nu}\xi^{i})_{\alpha}F_{\mu\nu}-D^{ij}\xi_{\alpha
j}\;,\\ &&\cr \delta_{\xi} \bar{\psi}^{
\dot{\alpha}i}&=&-\surd{2}( \bar{\sigma}^{\mu}\xi^{i})^{
\dot{\alpha}}\bigg{(}1+\frac{1}{\surd{2}}C_{s}
\bar{\phi}\bigg{)}\partial _{\mu} \bar{\phi}\;, \cr  &&\cr
\delta_{\xi}D^{kl}&=&\bigg{\{}-i\xi^{k} \sigma^{\mu}\partial
_{\mu}\bar{\psi}^{l}-i\xi^{l} \sigma^{\mu}\partial
_{\mu}\bar{\psi}^{k}\bigg{\}}\bigg{(}1+\frac{1}{\surd{2}}C_{s}
\bar{\phi}\bigg{)}. \nonumber\eea

It is straightforward to find a series of field redefinitions
which bring these deformed supersymmetry transformations to the
standard form. We introduce the multiplet $(a_{\mu},\varphi,
\bar{\varphi},\lambda_{\\alpha}^{i},\bar{\lambda}^{
\dot{\alpha}i},\tilde{D}^{ij})$ as \bea\label{drv5} a_{\mu}&=&F(
\bar{\phi})A_{\mu}\;,\cr  &&\cr \varphi&=&F(
\bar{\phi})^{2}\bigg{(}\phi+C_{s}\frac{1}{2\surd{2}}A_{\mu}A^{\mu}\bigg{)}\;,\cr
&&\cr \bar{\varphi}&=&\bar{\phi}\;,\cr  &&\cr \bar{\lambda}^{
\dot{\alpha}i}&=&F( \bar{\phi}) \bar{\psi}^{ \dot{\alpha}i}\\
&&\cr \lambda_{\alpha}^{i}&=&F(
\bar{\phi})^{2}\psi_{\alpha}^{i}\;,\cr  &&\cr \tilde{D}^{ij}&=&F(
\bar{\phi})^{2}D^{ij},\nonumber \eea
 where $F( \bar{\phi})$ is a function of $
\bar{\phi}$ and is determined as
\begin{eqnarray*}
F( \bar{\phi})=\frac{1}{1+\frac{1}{\surd{2}}C_{s} \bar{\phi}}.
\end{eqnarray*}

It is easy to check that the multiplet $(a_{\mu},\varphi,
\bar{\varphi},\lambda_{\\alpha}^{i},\bar{\lambda}^{
\dot{\alpha}i},\tilde{D}^{ij})$ transforms canonically under
supersymmetry transformations \bea
\delta_{\xi}a_{\mu}&=&i\xi^{i}\sigma_{\mu} \bar{\lambda}_{i},\cr
&&\cr \delta_{\xi}\varphi&=&-i\surd{2}\xi^{i}\lambda_{i},\cr &&\cr
\delta_{\xi}\bar{\varphi}&=&0,\cr  &&\cr
\delta_{\xi}\lambda_{\alpha}^{i}&=&(\sigma^{\mu\nu}\xi^{i})f_{\mu\nu}-
\tilde{D}^{ij}\xi_{\alpha j},\\  &&\cr
\delta_{\xi}\bar{\lambda}^{ \dot{\alpha}i
}&=&-\surd{2}(\bar{\sigma}^{\mu}\xi^{i})^{\dot{\alpha}}\partial
_{\mu} \bar{\varphi},\cr  &&\cr
\delta_{\xi}\tilde{D}^{ij}&=&-i\bigg{(}\xi^{i}\sigma^{\mu}\partial
_{\mu} \bar{\lambda}^{j}+\xi^{j}\sigma^{\mu}\partial _{\mu}
\bar{\lambda}^{i}\bigg{)}. \nonumber\eea

Here $f_{\mu\nu}=\partial _{\mu}a_{\nu}-\partial _{\nu}a_{\mu}$ .
 In the singlet deformed $U(1)$ theory which we consider here the action can be
constructed as \cite{Ferrara:2003xk} \bea S=\frac{1}{4}\int
d^{4}x_{L} d^{4}\theta du W*W =\frac{1}{4}\int d^{4}x_{L}
d^{4}\theta du W^{2}, \eea
 where $x_{L}$ is a chiral-analytic
coordinate:
\begin{eqnarray*}
x_{A}^{\mu}=x_{L}^{\mu}-2i\theta ^{-} \sigma^{\mu}
\bar{\theta}^{+}.
\end{eqnarray*}

To compute the filed strength $W=\frac{1}{4}
(\bar{D}^{+})^{2}V^{--}$ one can obtain $V^{--}$ from this
differential equation \bea
D^{++}V^{--}-D^{--}V^{++}+i[V^{++},V^{--}]_{*}=0 ,\eea
 here
\begin{eqnarray*}
D^{--}=\partial ^{--}-2i\theta^{-}\sigma^{\mu}
\bar{\theta}^{-}\frac{\partial}{\partial x_{A}^{\mu}}.
\end{eqnarray*}
In terms of the deformed analytic superfield  (\ref{drv0}), $W$ is
 calculated as \bea
W&=&\bigg{[}\phi+\frac{C_{s}}{2\surd{2}}A_{\mu}A^{\mu}\bigg{]}\cr
&&\cr
&+&2\theta^{+}[\psi^{i}u_{i}^{-}]-\frac{2}{1+\frac{1}{\surd{2}C_{s}
\bar{\phi}}} \theta^{-}[\psi^{i}u_{i}^{+}]\cr&&\cr
&+&(\theta^{+})^{2}\bigg{[}\frac{2C_{s}}{1+\frac{1}{\surd{2}C_{s}
\bar{\phi}}} \bar{\psi}^{i}
\bar{\psi}^{j}+D^{ij}\bigg{]}u_{i}^{-}u_{j}^{-}\cr  &&\cr
&+&\bigg{(}\frac{1}{1+\frac{1}{\surd{2}C_{s}
\bar{\phi}}}\bigg{)}(\theta^{-})^{2}\bigg{[}\frac{2C_{s}}{1+\frac{1}{\surd{2}C_{s}
\bar{\phi}}} \bar{\psi}^{i}
\bar{\psi}^{j}+D^{ij}\bigg{]}u_{i}^{+}u_{j}^{+}\\  &&\cr
&-&\bigg{(}\frac{2}{1+\frac{1}{\surd{2}C_{s}
\bar{\phi}}}\bigg{)}(\theta^{+}\theta^{-})\bigg{[}\frac{2C_{s}}{1+\frac{1}{\surd{2}C_{s}
\bar{\phi}}} \bar{\psi}^{i}
\bar{\psi}^{j}+D^{ij}\bigg{]}u_{i}^{-}u_{j}^{+}\cr  &&\cr
&+&(\theta^{+}\sigma^{\mu\nu}\theta^{-})F_{\mu\nu}
+2i(\theta^{-})^{2}\theta^{+}\sigma^{\mu}\partial
_{\mu}\bigg{(}\frac{1}{1+\frac{1}{\surd{2}C_{s} \bar{\phi}}}
\bar{\psi}^{i}\bigg{)}u_{i}^{+}\cr  &&\cr&+&
2i(\theta^{+})^{2}\theta^{-}\bigg{(}1+\frac{1}{\surd{2}C_{s}
\bar{\phi}}\bigg{)}\sigma^{\mu}\partial
_{\mu}\bigg{(}\frac{1}{1+\frac{1}{\surd{2}C_{s} \bar{\phi}}}
\bar{\psi}^{i}\bigg{)}u_{i}^{-}
-(\theta^{+})^{2}(\theta^{-})^{2}\partial ^{2}
\bar{\phi},\nonumber \eea

If we use redefined fields (\ref{drv5}) to compute the Lagrangian
 we will see that it has this simple form
\bea L=\bigg{(}1+\frac{1}{\surd{2}C_{s}
\bar{\phi}}\bigg{)}^{2}L_{0}, \eea
 where
\begin{eqnarray*}
L_{0}&=&\frac{1}{4}f_{\mu\nu}(f^{\mu\nu}+
\tilde{f}^{\mu\nu})-i\lambda ^{i}\sigma ^{\mu}\partial
_{\mu}\bar{\lambda}_{i}-\partial ^{\mu}\varphi \partial _{\mu}
\bar{\varphi}+\frac{1}{4}\tilde{D}_{ij}\tilde{D}^{ij}.
\end{eqnarray*}

In this paper, using the Seiberg-Witten map, we have determined
the generalized analytic superfield and gauge parameter of ${\cal
N}=2$ supersymmetric $U(1)$ gauge theory to the
non(anti)commutative harmonic superspace for which the component
fields transform canonically under gauge transformations. The
component fields are then redefined to preserve the standard form
of supersymmetry transformations. With this redefined component
fields, the Lagrangian is obtained which has the same form as the
one in \cite{Ferrara:2004zv}.

\section*{Acknowledgments}

I would like to thank M. Alishahiha and A. E. Mosaffa for useful
comments and discussions and M. Abolhasani for his sincere
support.

\end{document}